\documentclass{icrc2009}

\usepackage{graphicx}   
\usepackage[caption=false]{caption}    
\usepackage[font=footnotesize]{subfig} 
\usepackage{fixltx2e}
\usepackage{url}

\newcommand{\shorttitle}[1]%
{\markboth{Proceedings of the 31\MakeLowercase{$^{st}$} ICRC, {\L}\'{o}d\'{z} 2009}{#1} }
\newcommand{\etal}{\MakeLowercase{\textit{et al. }}} 

\begin{document}
\title{Correlation between the synchrotron and the inverse-Compton components, in the longterm light curve of GeV blazars}
\author{\IEEEauthorblockN{Bagmeet Behera\IEEEauthorrefmark{1}\IEEEauthorrefmark{2},
			  Marcus Hauser\IEEEauthorrefmark{1}, and
			  Stefan J. Wagner\IEEEauthorrefmark{1}}
                            \\
\IEEEauthorblockA{\IEEEauthorrefmark{1}Landessternwarte, Zentrum f\"ur Astronomie der Universit\"at Heidelberg, D-69117, Germany}
\IEEEauthorblockA{\IEEEauthorrefmark{2}Fellow of the International Max Planck Research School for Astronomy and Cosmic Physics\\
at the University of Heidelberg}
}

\shorttitle{Behera \etal Optical-GeV correlations in blazars}
\maketitle

\begin{abstract}
To explore the correlation in the GeV band (i.e. the inverse-Compton component) with the optical band (i.e. the synchrotron component) of blazars that are bright in high energies, the light curves of a number of blazars monitored with the Fermi Gamma-ray Space Telescope (FGST), were compared to the corresponding light curves recorded with the Automatic Telescope for Optical Monitoring (ATOM). The results are presented.\\
\end{abstract}

 
\section{Introduction}
Blazars are an enigma of nature. Observations in all wavelengths from radio to TeV $\gamma$-rays are being done to get a clear understanding of the physical phenomena that produces the observed emission. Many advances have been made in this field and the lower energy component from radio up to X-rays can be explained using a synchrotron emitting particle population. The high energy component is observed using various $\gamma$-ray instruments; satellite borne instruments are used to measure fluxes up to $\approx$ a few $100$\,GeV and ground based Cherenkov telescopes are used for measurements in the VHE regime. The physical phenomena causing this high energy component is not unambiguously explained yet. Among the possible mechanism, the synchrotron-inverse-Compton models can explain the emission in many blazars, connecting the low energy component with the high energy part. These have however some deficiencies in terms of a wide range of parameters that are not very well constrained and flares that are still not clearly understood. These models are not dynamic while being self-consistent at the same time. There are also hadronic models, which have their own deficiencies. To test these models and derive set of parameters that can adequately explain the dynamic range of spectral properties in both the high and low energy regime, simultaneous observations are essential.\\

With the launch of the FGST (Fermi Gamma-ray Space Telescope) it is now possible to measure and monitor the $100$\,MeV to $\sim 100$\,GeV fluxes (and spectral indices, at least for the bright sources) of blazars. These measurements would be unbiased towards flaring states, due to FGSTs high sensitivity. In the optical band, with the blazar-monitoring program using the ATOM instrument in Namibia, simultaneous long-term light curves of many of these blazars are being recorded. This gives us the first opportunity where longterm correlations between the synchrotron and inverse-Compton components of multiple blazars can be studied that are unbiased towards bright flaring states in either wavelength regimes. We take a few selected blazars that have simultaneous Fermi (from publicly available data) and ATOM light curves to make such a correlation study. The resulting light curves show a wide range of variability in both bands, as well as different degrees of correlations for the various sources. The results of these correlation studies are presented.\\

\section{Instruments and Data}
The \textbf{A}utomatic \textbf{T}elescope for \textbf{O}ptical
\textbf{M}onitoring (ATOM) is a $1$\,m class telescope located near the
H.E.S.S. IACT array and is operated by the
H.E.S.S. collaboration. It is used for the automatic optical
monitoring of known and potential TeV emitting AGN, as well as
other sources of interest for the H.E.S.S. collaboration.  On a
typical night, we observe around 40 different targets with
multi-color imaging.\\
  \begin{figure*}[!th]
   \centerline{\subfloat[]{\includegraphics[width=2.75in]{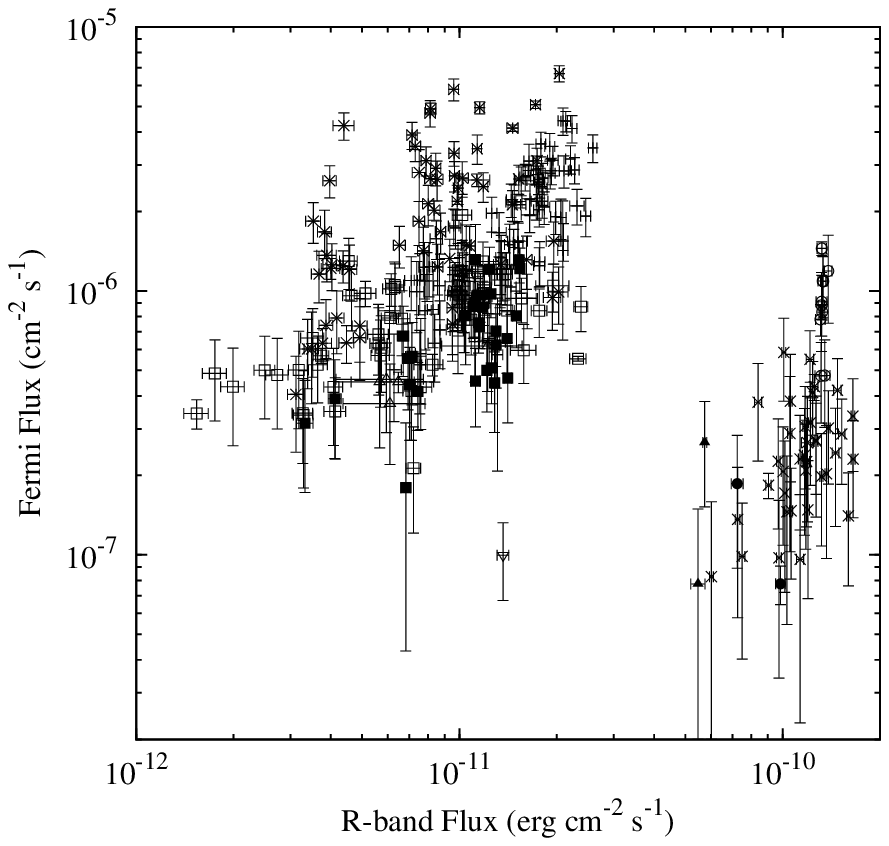}
	  \label{RbandVrsGeV}}
              \hfil
              \subfloat[]{\includegraphics[width=2.75in]{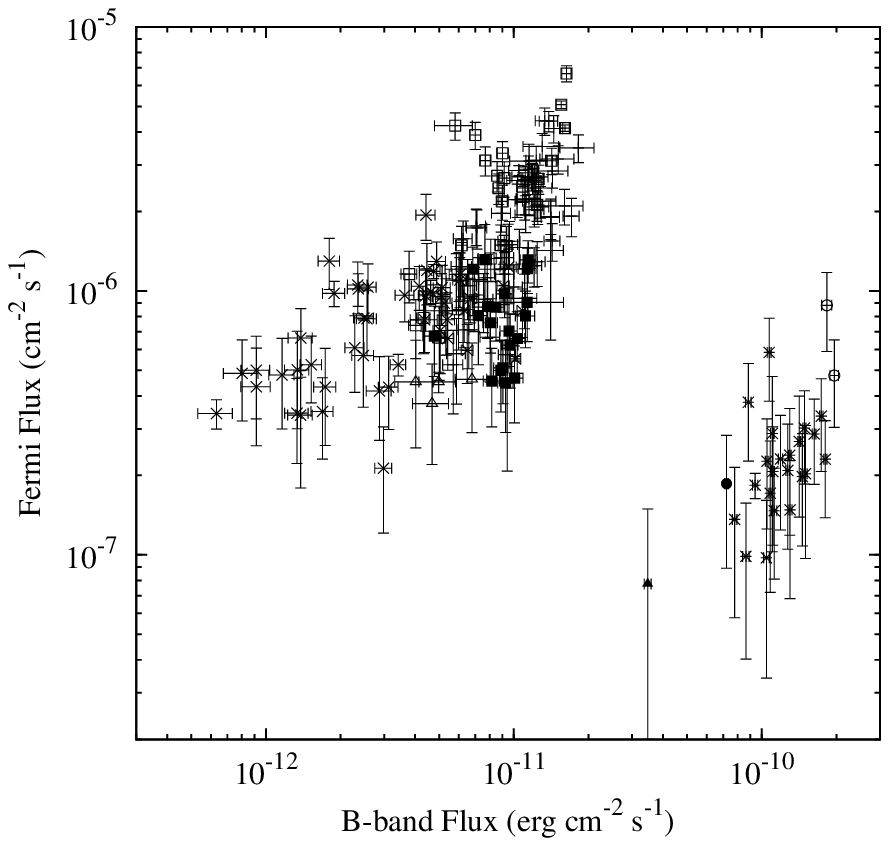}
	  \label{BbandVrsGeV}}
             }
   \caption{\textbf{\emph{left:}} The Fermi flux (from 0.1 GeV to 300 GeV) compared to the ATOM optical flux in the R-band. 10 sources listed in the text are plotted. Not all sources have equal number of points. A clear division into 2 groups is seen in the flux-flux regime. \textbf{\emph{right:}} The  Fermi flux compared to the corresponding ATOM B-band optical flux for 9 sources. The same pattern is seen as that in the left panel.}
   \label{FluxFlux}
\end{figure*}

The telescope is operated in a completely automatic mode, which means the
telescope system gets a nightly schedule via Internet, prepared before the start of nightly observations. It
then performs all the observations during the entire night without
any human interaction or supervision.\\

Data analysis is also done automatically by an analysis pipeline run on
site in the morning. This pipeline does the standard image
reduction (i.e. de-biasing and flat fielding) as well as finding an
astrometric solution for each frame down to arc second accuracy using
\emph{wcstools} \cite{Mink2005} and the \emph{UCAC2}
catalog \cite{Zacharias2004}. The
subsequent source detection and photometry is based on the
\emph{SExtractor} package \cite{Bertin1996}. The resulting source catalog
from each frame is then combined with older observations of the same
target and archived.\\

Since August 2008, FGST is operating in an all-sky scanning mode.
The Large Area Telescope (LAT) is one of the instruments on board the FGST
satellite. It has a large field of view, and detects gamma-rays in the range
of 20 MeV to 300 GeV. The FGST team provides daily flux measurements in
multiple FGST energy-bands on a list of 23 blazars, called the ``LAT Monitored
source'' which is publicly available at -  (\url{http://fermi.gsfc.nasa.gov/ssc/data/policy/LAT_Monitored_Sources.html}).
 Data is updated on roughly a daily basis. The fluxes are available in bins of $0.3$\,GeV to $1.0$\,GeV, $1.0$\,GeV to $300$\,GeV,
and  $0.1$\,GeV to $100.0$\,GeV. Spectral information is not provided in these
tables. Upper limits in these tables are ignored and only the measured flux
points are used for this correlation study.\\
\begin{figure*}[!th]
   \centerline{\subfloat[]{\includegraphics[width=2.75in]{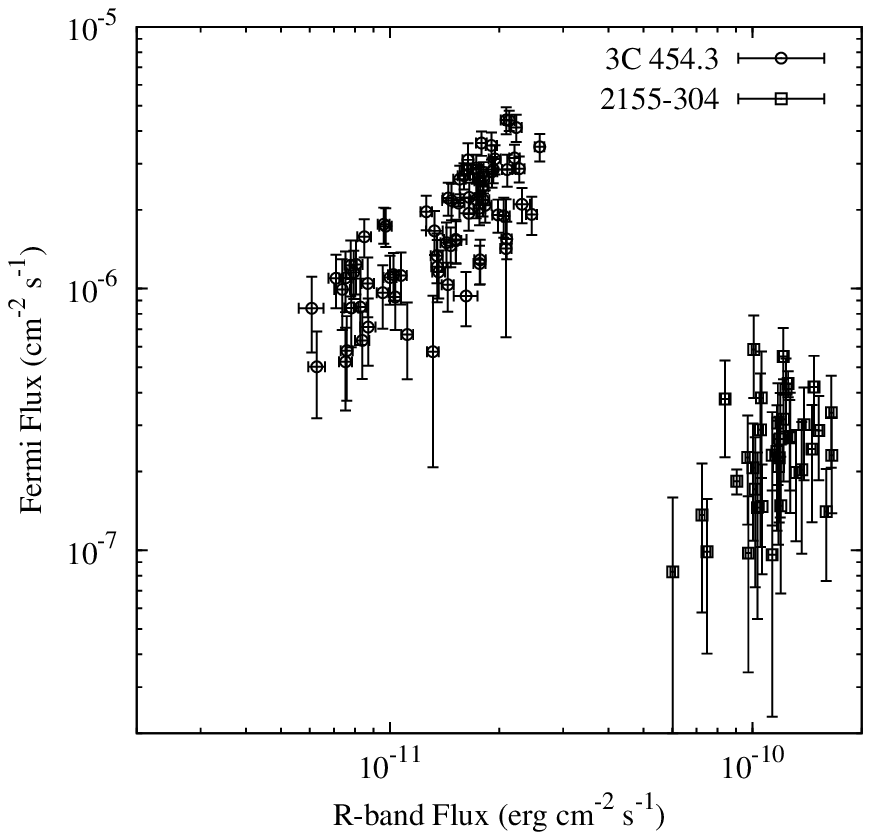}
	  \label{3C454.3}}
              \hfil
              \subfloat[]{\includegraphics[width=2.75in]{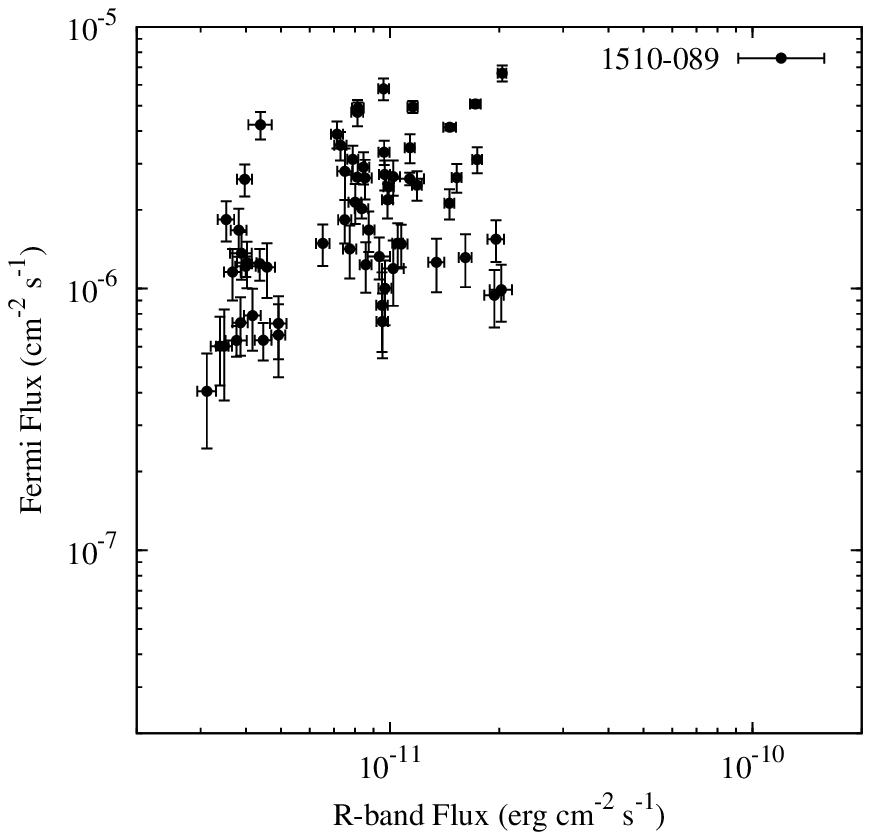}
	  \label{PKS1510-089}}
             }
   \caption{The Fermi flux versus the ATOM R-band flux for individual sources. \textbf{\emph{left:}} 3C\,454.3 and PKS\,2155-304, clearly showing the dichotomy among the two GeV-flux versus optical-flux regimes (see text), and within each regime a clear GeV-optical flux correlation is seen. \textbf{\emph{right:}} PKS\,1510-089 is the counter-example which shows a wide spread in the GeV-flux versus optical-flux plot.}
   \label{IndivSrcsFluxFlux}
\end{figure*}
\begin{figure*}[!th]
   \centerline{\subfloat[]{\includegraphics[width=2.75in]{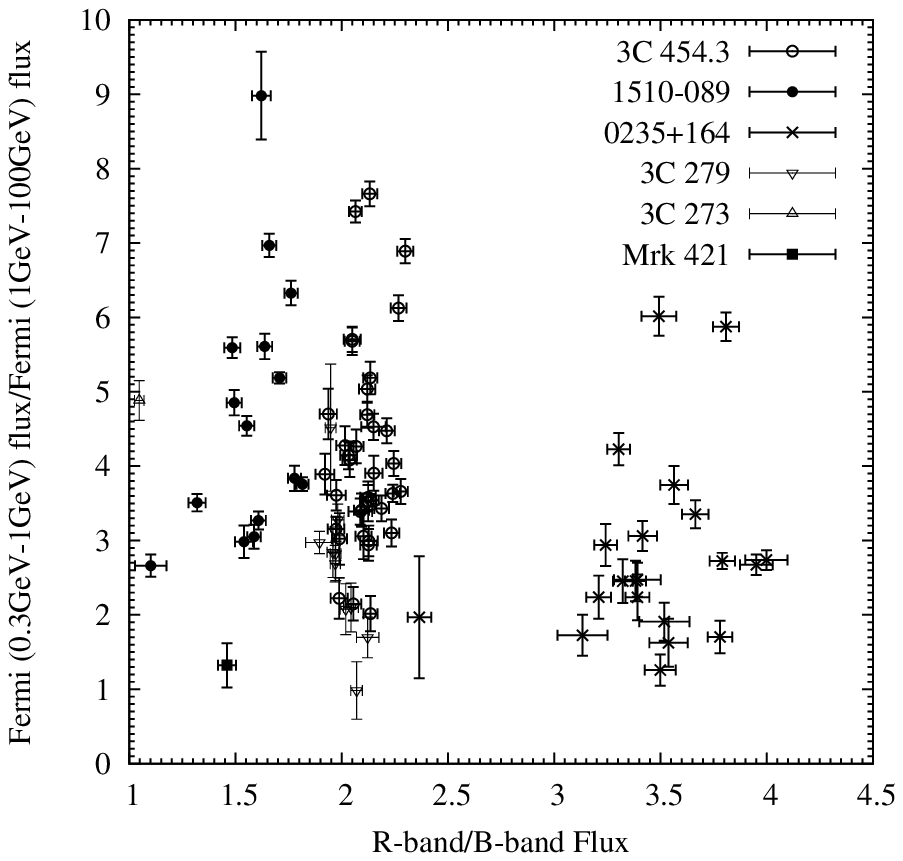}
	  \label{GeVspectraVersesOptSpec}}
              \hfil
              \subfloat[]{\includegraphics[width=2.75in]{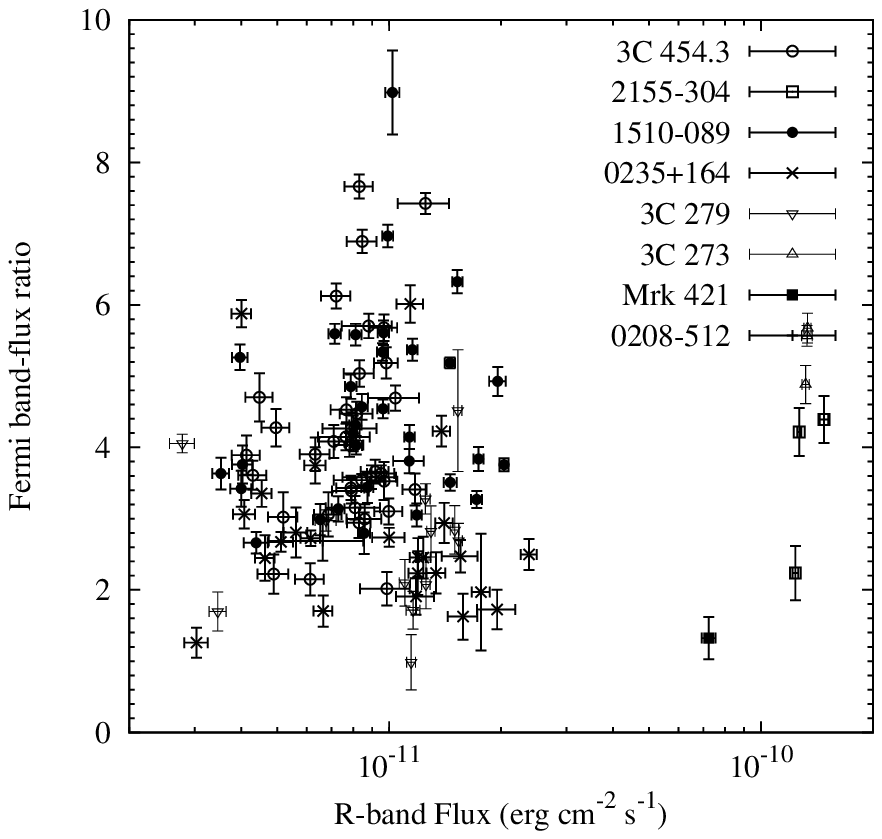}
	  \label{GeVspectraVersesOptFlux}}
             }
   \caption{\textbf{\emph{left:}} Ratio of the Fermi band-fluxes (low energy band-flux over high) is compared to the ratio of ATOM R-band flux to B-band flux \textbf{\emph{right:}} Ratio of the Fermi bands (low energy band-flux over high) compared to the ATOM R-band flux.}
   \label{SpectraSpectra}
\end{figure*}

\section{Results and Discussion} 
FGST data as obtained from the public server is in daily-averaged bins. A search is made on all the ATOM data in the R-band and B-band to find all measurements simultaneous with the FGST data points. Since FGST gives daily fluxes integrated over a $24$\,hour period (midnight to midnight, in UTC), and ATOM exposures are typically of the order of few minutes; all ATOM exposures within a particular $24$\,hour overlapping period are averaged to get a corresponding daily-average. Thus the light curves and data points we consider here are insensitive to intra-day variabilities in either band. Flux averaged over longer periods are not considered in this work.\\

The resulting simultaneous R-band optical and FGST flux measurements for $10$ sources are shown in figure \ref{RbandVrsGeV}. The ten sources are 3C\,454.3, PKS\,1510-089, 0235+164, 3C\,279, 0208-512, 1633+382, OJ\,287, PKS\,2155-304, 3C\,273, Mrk\,421 and Mrk\,501. A clear dichotomy is seen in the distribution of the points in this plot. The last $4$ sources in the list above, are in the bottom right corner, corresponding to high R-band flux and low Fermi flux; the first $5$ in the list are in the top-left corner of the plot, corresponding to relatively low R-band flux and relatively high Fermi flux. For OJ\,287 we get the single isolated point roughly in the middle of the two cluster of points. For the simultaneous B-band and FGST measurements we get a total $9$ sources, same as the previous list (except OJ\,287), shown in figure \ref{BbandVrsGeV}. A similar dichotomy is seen as in the previous case. Within either of the sub-sets there seems to be a positive correlation in the GeV and optical fluxes. The only exception being PKS\,1510-089, which lacks a clear correlation. Figure \ref{3C454.3} shows the GeV flux versus the R-band flux for 3C\,454.3 and PKS\,2155-304, clearly showing the dichotomy as well as the correlation between GeV and optical fluxes; whereas figure \ref{PKS1510-089} shows PKS\,1510-089, which shows no correlation.\\

The observed dichotomy may be explained if we have a sample with two different sub-classes of blazars. A simpler explanation would be a high host-galaxy optical-flux that moves a number of sources toward the right of the plots. However a simultaneous lower GeV flux cannot be explained away easily. Also interesting is the fact that the redshift of the sources also follow this dichotomy. The optically-faint and GeV-bright blazars all have redshift greater than $0.36$ (that of PKS 1510-089), where as the optically-bright and GeV-faint blazars are all nearby with redshifts less than $0.158$ (that of 3C\,273). Once more OJ\,287 falls in the middle of these two regimes with a $z\approx0.3$. This presents an interesting scenario, a more thorough study will be presented else-where.\\

Comparison of the spectra in the GeV band and the optical wavelengths is done using the flux-ratios as the proxy for spectral-index. For Fermi energies, the ratio of the flux in the low-energy band ($0.3$\,GeV to $1$\,GeV) over the high energy band ($1$\,GeV to $100$\,GeV) is compared with the ratio of the R-band flux over B-band flux, shown in figure \ref{GeVspectraVersesOptSpec}. We find no clear correlation in the GeV and optical spectral index values. For individual sources, the optical flux-ratio do not scatter much, where as there is comparably much higher scatter in the Fermi flux-ratio (hence spectral-index). Since most of these sources are variable in both GeV and optical bands, we can say that - the GeV spectral index varies with the variations in the GeV-flux level, whereas the optical color does not change much with changes in the optical-flux. We give a caveat that the flux ratio is not linearly related to spectral-index, hence is only a rough indicator of the later.\\

\section{Conclusion}
For $10$ GeV-bright-blazars the GeV and optical spectra are compared. Interesting correlation of the GeV fluxes and optical fluxes is found in most. A roughly linear correlation is found in the log-log representation of the flux comparison (i.e. a power law correlation is seen) for all sources except PKS\,1510-089. This blazar shows a broader scatter than others. A clear dichotomy is observed which divides the GeV-optical flux plane into two parts. The optically-faint- GeV-bright blazars consisting of sources with comparatively high redshift ($z > 0.36$). Whereas the optically-bright-GeV-faint blazars consist of sources with comparatively low redshift ($z < 0.158$). An intermediate redshift source OJ\,287 is found straddling the region in between these regimes with a redshift of $z=0.3$. The sample contains only 10 sources, thus these numbers and features might change with a larger sample. No clear correlation in the GeV spectral-index\newpage \noindent with optical color is detected. The GeV-flux versus optical-flux correlation presents an intriguing scenario that needs further study.\\

\end{document}